\documentclass[
    ,final            
    ,numberedheadings 
    ,cmfonts          
  ]
  {aipproc}

\layoutstyle{6x9}

\usepackage{amssymb}
\usepackage{graphicx}
\usepackage{bm}
\usepackage{citesort}
\usepackage{makeidx}
\usepackage{multicol}
\usepackage[notref,notcite]{showkeys}
\usepackage{bbm}
\usepackage{amsmath}
\usepackage{epsfig}
\long\def\comment#1{ }

\newcommand{\beq}{\begin{eqnarray}}
\newcommand{\eeq}{\end{eqnarray}}
\newcommand{\be}{\begin{eqnarray}}
\newcommand{\ee}{\end{eqnarray}}
\newcommand{\E}{Eq.~\eqref}

\newcommand{\BQ}{\begin{equation}}
\newcommand{\EQ}{\end{equation}}
\newcommand{\BQA}{\begin{eqnarray}}
\newcommand{\EQA}{\end{eqnarray}}

\newcommand{\abar}{\bar{\alpha}_s}
\newcommand{\lan}{\langle}
\newcommand{\ran}{\rangle}

\newcommand{\mean}[1]{\left\langle #1 \right\rangle_\tau}

\newcommand{\dif}{{\rm d}}

\def\simge{\mathrel{%
   \rlap{\raise 0.511ex \hbox{$>$}}{\lower 0.511ex \hbox{$\sim$}}}}
\def\simle{\mathrel{
   \rlap{\raise 0.511ex \hbox{$<$}}{\lower 0.511ex \hbox{$\sim$}}}}
\def\bigs{\mathrel{
   \rlap{\raise 0.531ex \hbox{$>$}}{\lower 0.531ex \hbox{$<$}}}}

\def\empile#1\over#2{\mathrel{\mathop{\kern 0pt#1}\limits_{#2}}}

\def\del{\partial}                              

\newcommand{\rme}{{\rm e}}

\newcommand{\x}{\bm x}
\newcommand{\y}{\bm y}

\newcommand{\z}{\bm z}

\renewcommand{\theequation}{\thesection.\arabic{equation}}

\begin{document}

\title{Fluctuating pulled fronts \& Pomerons}




\author{Edmond Iancu\footnote{Membre du Centre National de
la Recherche Scientifique (CNRS), France.}\ }{
  address={Service de Physique Th\'eorique, CEA/DSM/SPhT,  Unit\'e de recherche
associ\'ee au CNRS,\\ CE Saclay,
        F-91191 Gif-sur-Yvette, France} }

\begin{abstract}

I give a physical discussion of the influence of particle number
fluctuations on the high energy evolution in QCD. I emphasize the
event--by--event description and the correspondence with the problem of
`fluctuating pulled fronts' in statistical physics. I show that the
correlations generated by fluctuations reduce the phase--space for BFKL
evolution up to saturation. Because of that, the evolution `slows
down', and the rate for the energy increase of the saturation momentum
is considerably decreased. Also, the stochastic aspects inherent in
fluctuations lead to the breakdown of geometric scaling and of the BFKL
approximation. Finally, I explain the diagrammatic interpretation of
the particle number fluctuations as initiators of the Pomeron loops.

\end{abstract}

\maketitle


\section{Introduction}

Much of the recent progress in our understanding of QCD evolution at
high energy has been triggered by the observations that \texttt{(i)}
the gluon number fluctuations play an important role in the evolution
towards saturation and the unitarity limit \cite{MS04,IMM04} and
\texttt{(ii)} the QCD evolution in the presence of fluctuations and
saturation is in the same universality class as a series of problems in
statistical physics, the prototype of which being the
`reaction--diffusion' problem \cite{MP03,IMM04,IT04}.

These observations have developed into a profound and extremely
fruitful correspondence between high--energy QCD and modern problems in
statistical physics, which relates topics of current research in both
fields, and which has already allowed us to deduce some insightful
results in QCD by properly translating the corresponding results from
statistical physics \cite{IMM04,IT04}.

At the same time, the recognition of the importance of fluctuations has
revived the interest in the dilute regime of QCD at high energy, which
has been somehow overlooked by the modern theory for gluon saturation,
the Color Glass Condensate (CGC) \cite{MV,RGE,CGCreviews}. As first
noticed in Ref. \cite{IT04}, the evolution equation for the CGC
effective theory (also known as the JIMWLK equation \cite{JKLW,RGE,W})
does not include the `gluon splittings' responsible for gluon number
fluctuations (see Sect. 6 below), and the same is true also for the
Balitsky equations \cite{B} which describe the equivalent evolution of
the scattering amplitudes. On the other hand, the particle number
fluctuations are correctly taken into account (in the limit where the
number of colors $N_c$ is large) by Mueller's `color dipole' picture
\cite{Mueller}, and in fact it was within that context that Salam has
first observed, through numerical simulations \cite{Salam}, the
dramatic role played by fluctuations in the course of the evolution.

Thus, not surprisingly, the dipole picture occupies a central role in
the recent developments aiming at the inclusion of the effects of
particle number fluctuations in the non--linear evolution towards
saturation \cite{IT04,MSW05,IT05,LL05,KL05,BIIT05,Levin05}. 
Furthermore, the dipole
picture will also play a crucial role in the presentation that we shall
give here, and which is largely adapted from Refs.
\cite{MS04,IMM04,IT04,IT05}.

\section{The Balitsky--Kovchegov equation}
\setcounter{equation}{0}

The simplest physical context in which one can address the study of
gluon saturation is the collision between a small {\em color dipole} (a
quark--antiquark pair in a colorless state) and a high energy hadron
(the ``target''). At high energy, the target wavefunction is dominated
by gluons, to which couple the quark and the antiquark in the dipole.
Thus, by following the evolution of the dipole scattering amplitude
towards the unitarity limit, one can obtain information about the
evolution of the gluon distribution in the target towards saturation.

Since the projectile has such a simple structure, it is quite easy to
deduce the equation describing the evolution of the corresponding
$S$--matrix with increasing energy. We shall denote the $S$--matrix
element by $\langle S(\x,\y) \rangle_\tau$, where $\x$ and $\y$ are the
transverse coordinates of the quark and the antiquark, respectively,
and $\tau\sim \ln s$ is the `rapidity' variable, with $s$ the total
invariant energy squared. As we shall see, $\tau$ plays the role of an
`evolution time' for the quantum evolution with increasing energy. Now
suppose we increase $\tau$ by a small amount $d\tau$. In order to
compute the corresponding change in $\langle S\rangle_\tau$, it is more
convenient to keep the rapidity of the target fixed and put the small
change of rapidity into the elementary dipole. The latter then
`evolves', that is, it has a small probability of emitting a gluon due
to this change of rapidity, which can be estimated as
  \be \label{dPsplit} dP\,=\,
 \frac{\alpha_s N_c}{2\pi^2}\,{\mathcal M}({\bm{x}},{\bm{y}},{\bm z})\,
 {d^2\z}\,d\tau\,,\qquad {\mathcal M}({\bm{x}},{\bm{y}},{\bm z})\,\equiv\,
 \frac{(\x-\y)^2}{(\x-\z)^2(\y-\z)^2 }\,
  ,\ee
where $N_c$ is the number of colors and $\z$ is the transverse
coordinate of the emitted gluon. In the large--$N_c$ limit, to which we
shall restrict in what follows, the gluon can be effectively replaced by a
zero--size $q\bar q$ pair, and the gluon emission appears as the
splitting of the original dipole $(\x,\y)$ into two new dipoles
$(\x,\z)$ and $(\z,\y)$.

\begin{figure}[t]
    \centerline{\epsfxsize=4cm\epsfbox{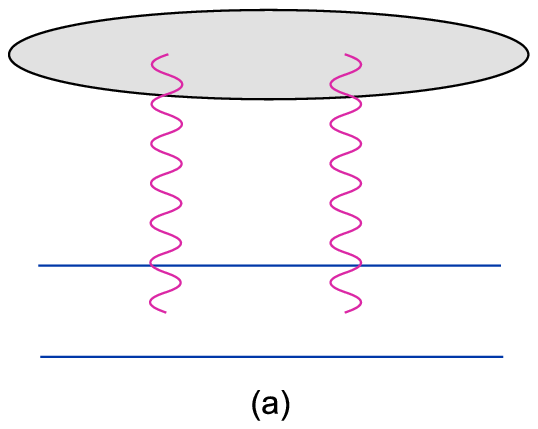}
    \epsfxsize=4cm\epsfbox{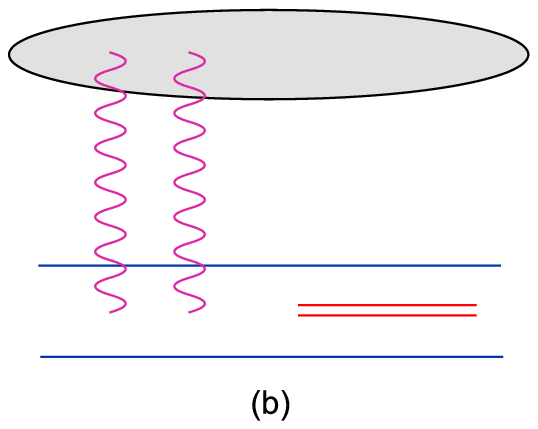}
    \epsfxsize=4cm\epsfbox{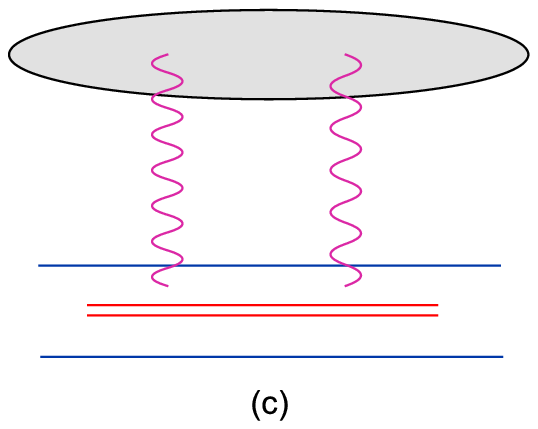}
    \epsfxsize=4cm\epsfbox{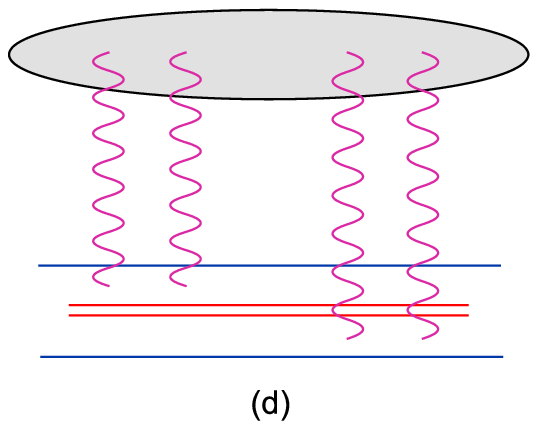}}
    \caption{\sl  Diagrams for the evolution of the dipole
    scattering amplitude, cf. \E{B1}: (a) the tree--level contribution;
    (b) the virtual correction $- \langle T(\x,\y)\rangle$;
    (c) the scattering of one child dipole, $\langle T(\x,\z)
    \rangle$ or $ \langle T(\z,\y)\rangle$;
    (d) the simultaneous scattering of both child dipoles,
    $\langle T^{(2)}(\x,\z;\z,\y)\rangle$.
    \label{FIG_1DIP}}
    \end{figure}

If the emitted gluon is in the wavefunction of the dipole at the time
it scatters on the target, then what scatters is a system of two
dipoles. If the gluon is not in the wavefunction at the time of the
scattering, it can be viewed as the ``virtual'' term which decreases
the probability that the original quark--antiquark pair remain a simple
dipole, thus compensating the probability for the two--dipole state.
The whole process can be summarized into the following {\em evolution
equation}, which has been originally derived by Balitsky \cite{B}:
 \be \frac{\del}{\del \tau}\langle S(\x,\y)
 \rangle_\tau =\frac{\bar{\alpha}_s}{2\pi} \!\int_{\z}\,
{\mathcal M}({\bm{x}},{\bm{y}},{\bm z})\, \big\{ - \langle S(\x,\y)
 \rangle_\tau + \langle S^{(2)}(\x,\z;\z,\y)
 \rangle_\tau\big\},
 \label{BS} \ee
where $\bar{\alpha}_s = {\alpha}_s N_c/\pi$ and $\langle
S^{(2)}(\x,\z;\z,\y) \rangle_\tau$ stands for the scattering of the
two--dipole system on the target. For what follows, it is more useful
to rewrite this equation in terms of the {\em scattering amplitude}
$T=1-S$. (Indeed, we shall be mostly concerned with the weak scattering
regime where $S$ is close to one --- recall that $|S|^2$ represents the
probability that no interaction take place in the collision ---, and
thus $T$ is small: $T\ll 1$.) The corresponding equation reads:
 \be \label{B1}\frac{\del}{\del \tau}\langle T(\x,\y)
 \rangle_\tau &=&\frac{\bar{\alpha}_s}{2\pi} \!\int_{\z}\,
{\mathcal M}({\bm{x}},{\bm{y}},{\bm z})\, \\
 &{}&\big\{ - \langle T(\x,\y)
 \rangle_\tau + \langle T(\x,\z)
 \rangle_\tau + \langle T(\z,\y)
 \rangle_\tau - \langle T^{(2)}(\x,\z;\z,\y)
  \rangle_\tau\big\},\nonumber
  \ee
and is illustrated with a few Feynman graphs in Fig. \ref{FIG_1DIP}.
(For simplicity, in this figure we represent the scattering between
an elementary dipole and the target in the two--gluon exchange approximation.)

But although formally simple, \E{B1} is not a closed equation
--- it relates a single--dipole scattering amplitude
to a two--dipole one ---, and
the true difficulty refers to the evaluation of $\langle T^{(2)}
\rangle_\tau$. To that aim, we need some information about the target.
The simplest approximation is to assume factorization
 \be\label{fact}
 \langle T^{(2)}(\x,\z;\z,\y)\rangle_\tau
 \,\approx\,\langle T(\x,\z)\rangle_\tau\,\langle
 T(\z,\y)\rangle_\tau\,,\ee
which is a {\em mean field approximation} (MFA) for the gluon fields in
the target. This immediately yields a closed, non--linear, equation for
$\langle T \rangle_\tau$ :
 \be \frac{\del}{\del \tau}\langle T(\x,\y)
 \rangle_\tau &=&\frac{\bar{\alpha}_s}{2\pi} \!\int_{\z}\,
{\mathcal M}({\bm{x}},{\bm{y}},{\bm z})\, \\
 &{}&\big\{ - \langle T(\x,\y)
 \rangle_\tau + \langle T(\x,\z)
 \rangle_\tau + \langle T(\z,\y)
 \rangle_\tau - \langle T(\x,\z)\rangle_\tau\,\langle
 T(\z,\y)\rangle_\tau\big\}.\nonumber
 \label{BK} \ee
This is the equation originally derived by Kovchegov \cite{K}, and
commonly referred nowadays as the `Balitsky--Kovchegov (BK) equation'.
Remarkably, this equation predicts that the scattering amplitude should
approach the unitarity bound $T=1$ in the high energy limit. By
contrast, the linear version of this equation as obtained by neglecting
the term quadratic in $\langle T \rangle_\tau$ in its r.h.s. (this is
the celebrated BFKL equation \cite{BFKL}) predicts an exponential
growth of the amplitude with $\tau$, which would eventually violate
unitarity. But, of course, the linear approximation breaks down when
the average amplitude becomes of order one, since then the non--linear
term becomes important and restores unitarity. As manifest on Fig.
\ref{FIG_1DIP}, the non--linear effects reflect {\em multiple
scattering}.

The BK equation is perhaps the `best' simple equation for dealing with
the onset of unitarity in QCD at high energy and weak coupling.
However, what we are primarily interested here in are the {\em
limitations} of this equation, coming from the factorization assumption
\E{fact}. The latter may be a good approximation if the target is a
large nucleus and for not very high energies, which is the situation
for which Kovchegov has originally derived this equation. More
generally, this should work reasonably well when the scattering is
sufficiently strong, that is, when $\langle T \rangle_\tau$ is not much
smaller than one, because in that case the external dipole scatters off
a high--density gluonic system, and the {\em density fluctuations} are
relatively unimportant. On the other, the MFA cannot be right if the
scattering is {\em very weak}, because then the dipole is sensitive to
the {\em dilute} part of the target wavefunction, where the
fluctuations are, of course, essential. Still, given that our main
interest when using \E{BK} is in the {\em strong} scattering regime,
one may expect the limitations of this equation in the dilute
regime to be inessential for the problem at hand. However, this
expectation turns out to be incorrect, and this is precisely what we
would like to explain in what follows: The {\em particle
number fluctuations} in the dilute regime have a strong influence, via
their subsequent evolution, on the approach towards saturation and the
unitarity limit.

\section{The fate of the rare fluctuations }
\setcounter{equation}{0} \label{SECT_FLUCT}

Since the fluctuations are a priori important in the weak scattering
regime, we shall focus on the scattering of a {\em small dipole}, with
transverse size $r\equiv |\x-\y|\ll 1/Q_s(\tau)$. We have introduced
here the {\em saturation momentum} $Q_s(\tau)$ \cite{GLR}, which is a
characteristic scale of the gluon distribution in the target, and marks
the scale at which a dipole scattering off the target makes the
transition from weak ($r\ll 1/Q_s$) to strong ($r\gg 1/Q_s$)
interactions. It is in fact common to {\em define} $Q_s(\tau)$
by the condition 
 \be\label{Nsat} \langle T(\bm{x},\bm{y}) 
 \rangle_\tau\,=\,1/2\qquad{\rm for}\qquad
 r\,=\,1/Q_s(\tau)\,,\ee
and to use this condition together with the solution to the BK equation
\eqref{BK} in order to compute the energy dependence of the saturation
momentum. We shall discuss more about this in the next section.

Returning to our small external dipole, we would like to relate its
scattering amplitude to the average gluon density in the target. This
is indeed possible in the dilute regime, since then the dipole scatters
only once. In fact, at large $N_c$ we can achieve a more symmetric
description by representing also the gluons in the target as color
dipoles, with an {\em dipole number density} ${n(\bm{u},\bm{v})}$ (for
dipoles with a quark at $\bm{u}$ and an antiquark at $\bm{v}$). The
external dipole $(\bm{x},\bm{y})$ can scatter off any of the internal
dipoles $(\bm{u},\bm{v})$ by exchanging two gluons. This gives:
 \begin{align}\label{Tconv}
    \mean{T(\bm{x},\bm{y})} = \alpha_s^2
    \int \dif^2 \bm{u}\, \dif^2 \bm{v}\,
    {\cal{A}}_0(\bm{x},\bm{y}|\bm{u},\bm{v})\,
     \mean{n(\bm{u},\bm{v})},
 \end{align}
where $\alpha_s^2{\cal{A}}_0(\bm{x},\bm{y}|\bm{u},\bm{v})$ is the
scattering amplitude for two elementary dipoles. Here, we shall not
need its exact expression, but only the fact that it is {\em
quasi--local} both with respect to the dipole {\em sizes} and with
respect to their {\em impact parameters}. (The impact parameter of a
dipole $(\bm{x},\bm{y})$ is its center--of--mass coordinate $\bm{b}=
(\x+\y)/2$.) This allows us to simplify \E{Tconv} as:
 \be\label{Tf}
 \lan T(r,b)\ran_{\tau}
  \,\simeq \,\alpha_s^2 \,\lan f (r,b) \ran_{\tau}\,,\ee
where the dimensionless quantity
 \be\label{fdip}
   f(r,b)\,\simeq \,r^2
  \int_{\Sigma} d^2{\bm{b}'} \,n(\bm{r},\bm{b}')\ee
is the {\em dipole occupation number} in the target, that is, the
number of dipoles with size $r$ (per unit of $\ln r^2$) within an area
$\Sigma\sim r^2$ centered at $b$. Eq.~(\ref{Tf}) shows that a small
dipole projectile is a very precise analyzer of the dipole distribution
in the target: the external dipoles counts the numbers of internal
dipoles having the same transverse size and impact parameter as itself.

Eq.~(\ref{Tf}) applies so long as $\lan T\ran_{\tau}\ll 1$, but by
extrapolation it shows that unitarity corrections in the dipole--target
scattering become important when the dipole occupation factor in the
target becomes of order $1/\alpha_s^2$. This is precisely the critical
density at which {\em saturation effects} --- i.e., non--linear effects
in the target wavefunction leading to the saturation of the gluon
distribution --- are expected to occur \cite{Mueller}.
This argument confirms that, by
studying dipole scattering in the vicinity of the unitarity limit, one
has access at the physics of gluon saturation.

Let us assume an initial condition like (\ref{Tf}) at the initial
rapidity ${\tau_0}$ and follow the evolution of the scattering
amplitude with increasing $\tau$. At the beginning, the amplitude
will rise very
fast, according to the BFKL equation, but this rise will be eventually
stopped by the non--linear term $\lan T^{(2)}\ran_{\tau}\equiv \langle
T^{(2)}(\x,\z;\z,\y) \rangle_\tau$ in Eq.~(\ref{B1}), which in the
linear regime rises even faster. We have, schematically,
 \be\label{TT2}
 \lan T\ran_{\tau}\simeq T_{0}\,{\rm e}^{\omega_{\mathbb P}
 (\tau-\tau_0)}\,,\qquad \lan T^{(2)}\ran_{\tau} \simeq
 T^{(2)}_{0}\,{\rm e}^{2\omega_{\mathbb P} (\tau-\tau_0)}\,,\ee
where $\omega_{\mathbb P} = {\rm const.}\times \bar{\alpha}_s$, $T_0
\equiv \lan T\ran_{\tau_0}\simeq\alpha_s^2\, f_0$ and $T^{(2)}_{0}
\equiv \lan T^{(2)}\ran_{\tau_0}$. ($f_0$ denotes the {\em average}
occupation factor at $\tau={\tau_0}$.) The unitarity limit is approached
when $\lan T^{(2)}\ran_{\tau} \sim \lan T \ran_{\tau}$, which in turn
implies $\tau\sim \tau_c$ with
 \be\label{TAUC}
 \rme^{\omega_{\mathbb P} (\tau_c-\tau_0)}
    \,\sim\,{T_0}/{T^{(2)}_{0}}\,.\ee
So, what is the ratio $T^{(2)}_{0}/T_0$ ? If one
assumes the factorization property
(\ref{fact}), then $T^{(2)}_{0}\approx (T_0)^2$, and therefore
$T^{(2)}_{0}/T_0\approx T_0\simeq\alpha_s^2\, f_0$. Then
Eq.~(\ref{TAUC}) implies:
 \be\label{TAUC1}
 \tau_c-\tau_0
  \,\simeq\,\frac{1}{\omega_{\mathbb P}}
  \,\ln\frac{1}{\alpha_s^2 f_{0}}\,=\,\frac{1}{\omega_{\mathbb P}}
  \left(\ln\frac{1}{\alpha_s^2} +\ln\frac{1}{f_{0}}\right).\ee
But is the MFA (\ref{fact}) a reasonable approximation for a {\it
dilute} initial condition ? To answer this question, let us consider
two physical situations: \texttt{(i)} $f_0\gg 1$ (with $f_0\ll
1/\alpha_s^2$ though) and \texttt{(ii)} $f_0\ll 1$. Also, remember that
$ \langle T^{(2)}(\x,\z;\z,\y) \rangle_\tau$ is the scattering
amplitude for two incoming dipoles $(\x,\z)$ and $(\z,\y)$ which have
similar impact factors (since they have a common leg at $\z$) and also
similar sizes (since the QCD evolution, Eq.~(\ref{B1}), favors the
splitting into dipoles with similar sizes).

\texttt{(i)} In the first case, the disk $\Sigma\sim r^2$ at $b$ has a
high occupancy, so the two external dipoles will predominantly scatter
off {\em different} dipoles in that disk. Then, their scatterings are
largely independent, and the MFA is reasonable. The result
(\ref{TAUC1}) can thus be trusted in this case.

\texttt{(ii)} The statement that the {\em average}
occupation factor $f_0$ is much smaller than one requires an explanation.
Clearly, in a {\em given configuration} of the  target,
the occupation number (\ref{fdip}) is {\em
discrete} : $f=0,1,2,\dots$ ; so, for its {\em average} value to be
smaller than one, one needs to look at {\em rare configurations}. That
is, if one considers the {\em statistical ensemble} of dipole
configurations generated by the evolution up to rapidity $\tau_0$, then
for most of these configurations $f(r,b)=0$, but for a small fraction
among them, of order $f_0$, $f$ is non--zero and of order one. Thus,
$f_0$ is essentially the {\em probability} to find a dipole with the
required characteristics $(r,b)$ in the ensemble.

Consider now the scattering problem in such a {\em very} dilute regime:
The fact that $T_{0} \sim\alpha_s^2 f_0\ll \alpha_s^2$ means that the
incoming dipole $(r,b)$ has a small probability $f_0(r,b)$ to find a dipole
with similar characteristics in the target, with which it then
interacts with a strength $\alpha_s^2$. Consider now {\em two} incoming
dipoles, with similar sizes and impact parameters: there is a small
probability $f_0(r,b)$ to find a corresponding dipole in the target,
but whenever this happens, {\em both} external dipoles can scatter off
it, with an overall strength $\alpha_s^4$. This gives $T^{(2)}_{0} \sim
\alpha_s^4f_{0} \sim \alpha_s^2 T_0$, which is much larger than the MFA
prediction $T^{(2)}_{0} \sim  (T_0)^2$. The scattering of the external
dipoles is now {\em strongly correlated}. With this estimate for
$T^{(2)}_{0}$, Eq.~(\ref{TAUC}) implies
  \be\label{TAUC2}
 \tau_c-\tau_0
  \,\simeq\,\frac{1}{\omega_{\mathbb P}}
  \,\ln\frac{1}{\alpha_s^2}\,.\ee
For $f_0\ll 1$, this is considerably smaller than the naive estimate
(\ref{TAUC1}) based on the MFA. Thus, {\em by enhancing the
correlations in the dilute regime, the fluctuations in the particle
number significantly reduce the rapidity window for BFKL evolution.}

Moreover, at the rapidity $\tau_c$ at which the unitarity corrections
cut off the BFKL growth, Eqs.~(\ref{TT2}) and (\ref{TAUC2}) imply $\lan
T\ran_{\tau_c} \sim \lan T^{(2)} \ran_{\tau_c}\sim f_0 \ll 1$, in sharp
contrast with the prediction of the  MFA !
That is, the contribution that a {\em rare} fluctuation $(r,b)$ at
$\tau={\tau_0}$ can give,  through its subsequent evolution, to the
average amplitude $\lan T(r,b)\ran_{\tau}$ at $\tau >{\tau_0}$
saturates at a value {\em smaller} than one
(of the order of the probability
$f_0(r,b)\ll 1$ of the original fluctuation) \cite{MS04}. Besides, this
contribution violates the factorization assumption implicit in
the BK equation \cite{IT04}.

But then how can the average amplitude $\lan T(r,b)\ran_{\tau}$ ever
approach the unitarity limit $\lan T\ran_{\tau}= 1$ ? This is possible
because, as manifest on  Eq.~(\ref{B1}), the evolution is {\em
non--local} in $r$, that is, a dipole of size $r$ can split also from
dipoles of larger sizes $r'\gg r$, which in the original ensemble at
$\tau_0$ had a larger probability to exist, and thus an average
occupation factor $f_0(r',b)\ge 1$. At $\tau={\tau_0}$, these larger
dipoles were not `seen' by the external dipole $r$, because of the
mismatch in sizes, but their descendants of size $r$ at rapidity $\tau
>{\tau_0}$ {\em are} seen, and they actually dominate the scattering as
compared to the rare fluctuations discussed previously.

We conclude that the correlations in the dilute regime significantly
reduce the phase--space available for the BFKL evolution of the {\em average}
amplitude towards saturation, by eliminating the rare fluctuations
$(r,b)$ for which $\lan f(r,b)\ran_\tau < 1$, or, equivalently, for
which $\lan T(r,b)\ran_{\tau} < \alpha_s^2$ \cite{MS04}. The limiting
value $\alpha_s^2$ is the elementary {\em `quantum'} for the strength
of $T$ in the event--by--event description, that is, the minimal
non--trivial value that a physical scattering amplitude can take in a
particular event, where the dipole number is discrete \cite{IMM04}.

In view of this, one expects the evolution to `slow down'  as compared
to the MFA. This is confirmed by an original calculation by Mueller and
Shoshi \cite{MS04}, using a restricted BFKL evolution, which shows that
the rate for the growth of saturation momentum with the energy is
considerably reduced as compared to the corresponding prediction of the
BK equation. In the next section, we shall
recover the result of Ref. \cite{MS04} from a broader perspective,
which allows one to also study the statistical features of the
evolution towards saturation, through a remarkable correspondence with
modern results in statistical physics \cite{IMM04}.

\section{Fluctuating pulled fronts}
\setcounter{equation}{0} \label{SECT_EBE}

To perform a detailed study of the influence of fluctuations on the
evolution towards high density, one needs a theory for correlations
like $\lan T^{(2)} \ran_{\tau}$ in the presence of fluctuations. Such a
theory has been recently given (within the large--$N_c$ approximation)
\cite{IT04,IT05,MSW05}, and we shall briefly comment on it in the last
section. But before doing that, we would like to show that some very
general results concerning the effects of fluctuations can be deduced
without a detailed knowledge of the microscopic dynamics, by relying on
universal results from statistical physics  \cite{IMM04}.

Specifically, the only assumptions that we shall need in order to
derive these results are the following: \texttt{(i)} the {\em mean
field description} of the dynamics of $\lan T\ran_{\tau}$ is provided
by the BK equation (\ref{BK}), and \texttt{(ii)} in the {\em
event--by--event description}, the amplitude $T$ is a discrete
quantity, with step $\Delta T\sim \alpha_s^2$.

We start by summarizing those results about the BK equation that are
needed for the present purposes. We shall neglect the impact parameter
dependence of the amplitude, and write the corresponding solution as $\lan
T(r)\ran_{\tau}\equiv \overline {T}_\tau(\rho)$, where $\rho\equiv
\ln(r_0^2/r^2)$ and $r_0$ is a scale introduced by the initial
conditions at low energy. Note that small dipole sizes correspond to
large values of $\rho$. Thus, the amplitude is small, $\overline
{T}_\tau(\rho)\ll 1$, when  $\rho$ is sufficiently large: $\rho\gg
\bar\rho_s(\tau)$, where $\bar\rho_s(\tau)\equiv \ln(r_0^2 \bar
Q_s^2(\tau))$ and $\bar Q_s^2(\tau)$ denotes the saturation momentum
extracted from the BK equation.

\begin{figure}[t]    
    \centerline{\epsfxsize=15.cm\epsfbox{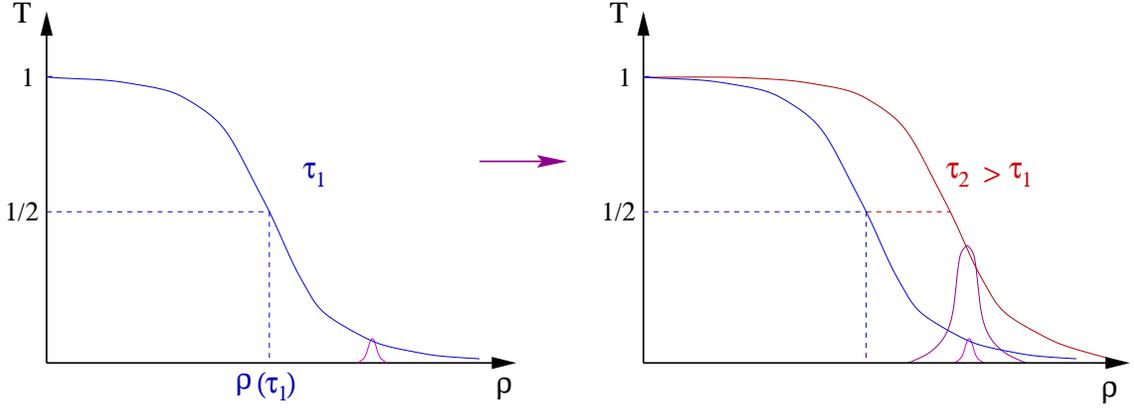}}
    \caption{\sl Evolution of the continuum front of the BK
    equation with increasing rapidity $\tau$.
                 \label{TWave5}}
                 \vspace*{0.5cm}
\end{figure}

The solution $\overline {T}_\tau(\rho)$ can be visualized as a {\em
front} which interpolates between $T=1$ (the unitarity limit)  at
$\rho\to -\infty$ and $T=0$ at $\rho\to \infty$ \cite{MP03} (see Fig.
\ref{TWave5}). Note that $T=1$ and $T=0$ are stable and, respectively,
unstable fixed points of the BK equation. The transition between the
two regimes occurs at $\rho\sim \bar\rho_s(\tau)$ ; thus, the
(logarithm of the) saturation momentum plays the role of the {\em
position of the front}. With increasing $\tau$, the saturation momentum
rises very fast (exponentially in $\tau$), so the front moves towards
larger values of $\rho$. One finds \cite{GLR,SCALING,MT02,DT02,MP03}:
  \be\label{Qsfix1}
Q_s^2(\tau)\,\simeq\,Q_0^2\, 
 \,\frac{
 {\rm e}^{c\bar\alpha_s \tau}}{(\bar\alpha_s \tau)^{3/2\gamma_s}} \,,\ee
where $Q_0^2\propto 1/r_0^2$, and $c$ and $\gamma_s$ are numbers fixed
by the BFKL dynamics: $c=4.88...$ and $\gamma_s=0.63..$. \E{Qsfix1}
implies the following expression for the {\em front velocity} :
 \be \label{lsdef}
 \bar\lambda(\tau)\,\equiv
   \,\frac{d\bar \rho_s(\tau)}
    {d\tau}\, \,\simeq \, c \bar\alpha_s \,-\, \frac{3}{2\gamma_s}\,
 \frac{1}{\tau}\,.\ee
Its asymptotic value at large $\tau$ represents the {\em saturation
exponent} (the rate for the exponential growth of $Q_s^2(\tau)$), here
estimated in the MFA: $\bar\lambda_{\rm as} = c\bar\alpha_s$.

In the weak scattering (dilute) regime at $\rho\gg \bar\rho_s(\tau)$,
the form of the amplitude can be obtained by solving the linearized
version of Eq.~(\ref{BK}), that is, the BFKL equation. One thus finds
(up to an overall normalization factor) \cite{SCALING,MT02,DT02,MP03}:
  \be\label{Trho}
 \overline {T}_\tau(\rho) \, \simeq \,(\rho-\bar\rho_s)\,
 {\rm e}^{-\gamma_s (\rho-\bar\rho_s)}\,
 \exp\left\{-\frac{(\rho-\bar\rho_s)^2}
 {2\beta \bar\alpha_s \tau} \right\},\ee
(with $\beta\simeq 48.2$). 
In particular, so long as the difference
$\rho-\bar\rho_s$ remains much smaller than the {\em diffusion radius}
$\sim \sqrt{2\beta \bar\alpha_s \tau}$, the Gaussian in
Eq.~(\ref{Trho}) can be ignored, and the amplitude becomes purely a
function of $\rho-\bar\rho_s(\tau)$ :
 \be\label{Trhos}\hspace*{-3mm}
 \overline {T}_\tau(\rho) \, \simeq \,\,(\rho-\bar\rho_s(\tau))\,
 {\rm e}^{-\gamma_s (\rho-\bar\rho_s(\tau))}\qquad{\rm for}\qquad
 \rho-\bar\rho_s\ll
 \sqrt{2\beta \bar\alpha_s \tau}\,.\ee
This is the property referred to as `geometric scaling'
\cite{geometric,SCALING}. It means that the front propagates
without distortion, as a {\em traveling wave} \cite{MP03}.

Notice the mechanism leading to the front propagation: For a fixed
$\rho\gg \bar\rho_s(\tau)$, the amplitude (\ref{Trho}) rises rapidly
with $\tau$, due to the exponential factor ${\rm exp}{(\gamma_s
\bar\rho_s)}\,\simeq\,{\rm e}^{\omega_{\mathbb P}\tau}$ with
$\omega_{\mathbb P}=\gamma_s \bar\lambda_{\rm as}\,$; this is the BFKL
instability (see Fig. \ref{TWave5}).
Thus the front is {\em pulled} by the unstable (BFKL) growth of its
tail at large $\rho$. Besides, for a given (large) distance
$\rho-\bar\rho_s$ ahead of the front, the amplitude increases through
{\em diffusion} from smaller values of $\rho$, until it reaches the
profile (\ref{Trhos}) of the traveling wave.

The fact that the front corresponding to the BK equation is a {\em
pulled front} --- it propagates via the growth and spreading of small
perturbations around the unstable state $T=0$ --- is crucial for the
problem at hand, as it shows that the front dynamics is driven by its
{\em leading edge} (the front region where $T\ll 1$), and therefore it
might be very sensitive to {\em fluctuations}. Although this property
has been discussed here on the basis of the linear, BFKL, equation, it
turns that this is an {\em exact} property of the non--linear BK
equation \cite{MP03}. Indeed, as shown by Munier and Peschanski, the BK
equation is in the same universality class as the
Fisher--Kolmogorov--Petrovsky--Piscounov (FKPP) equation \cite{FKPP},
which appears as a mean field approximation to a variety of stochastic
problems in chemistry, physics, and biology, and for which the pulled
front property has been rigorously demonstrated  (see \cite{Saar,Panja}
for recent reviews and more references).

Let us now return to the actual microscopic dynamics, which is {\em
stochastic} (it includes fluctuations in the number of dipoles in the
target), and where the scattering amplitude (in a given event) is {\em
discrete}. Then, as discussed in the previous section, one needs to
consider a {\em statistical ensemble of configurations}, which
correspond to different realizations of the same evolution. To any of
these configurations one can associate a front $T_\tau(\rho)$, which
characterizes the scattering between that particular configuration and
external dipoles of arbitrary size $\rho$.

\begin{figure}[t]    
    \centerline{\epsfxsize=15.cm\epsfbox{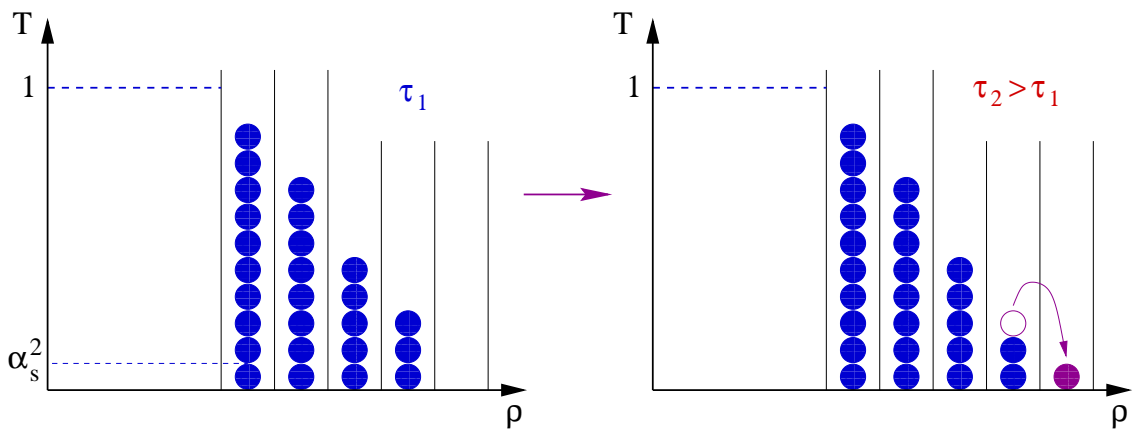}}
    \caption{\sl Evolution of the discrete front of a microscopic
    event with increasing rapidity $\tau$. The small blobs
    are meant to represent the elementary
    quanta $\alpha_s^2$ of $T$ in a microscopic event.
                 \label{TWave6}}
                 \vspace*{0.5cm}
\end{figure}

As in the mean field case, the evolution of a configuration is
described as the propagation of the associated front towards larger
values of $\rho$. What is however new is that, because of discreteness,
a microscopic front looks like a {\em histogram}: both $T_\tau$ and
$\rho$ are now discrete quantities, with steps $\Delta T=\alpha_s^2$
and $\Delta\rho = 1$, respectively. Because of that, the front is
necessarily {\em compact}
--- for any $\tau$, there is only a finite number of bins in $\rho$
ahead of $\rho_s$ where $T_\tau$ is non--zero (see Fig. \ref{TWave6})
 ---, and this property turns out to
have dramatic consequences for the propagation of the front:

In the empty bins on the right of the front tip, the local,
BFKL, growth is not possible anymore (this would require a seed !).
Thus, the only way for the front to progress there is via {\it
diffusion}, i.e., via radiation from the occupied bins at $\rho
<\rho_{\rm tip}$ (see Fig. \ref{TWave6}).
But since diffusion is less effective than the local
growth, we expect the velocity of the front --- which is also the
saturation exponent --- to be reduced for the microscopic front as
compared to the front of the MFA. The difference between the mechanisms
for front propagation in the MFA and in a microscopic event can be also
appreciated by comparing Figs. \ref{TWave5} and \ref{TWave6}.

The extreme sensitivity of the pulled fronts to small fluctuations has
been recognized in the context of statistical physics only in the
recent years, first via numerical simulations for discrete particle
models, then also through analytic arguments and physical
considerations \cite{BD}, that we have adapted to QCD in \cite{IMM04}.
The discrete particle version of a pulled front is generally referred
to as a {\em ``fluctuating pulled front''} \cite{Saar,Panja}. The most
striking feature of such a system is that the convergence towards the
mean field limit is extremely slow, {\em logarithmic} in the maximal
occupation number $N$. (For QCD, $N\sim 1/\alpha_s^2$, as explained
after Eq.~(\ref{fdip}).) Specifically, if $\lambda_N$ denotes  the
(asymptotic) velocity of the microscopic front for a finite value of
$N$, and $\lambda_\infty$ is the respective velocity in the MFA (which
corresponds to the limit $N\to\infty$), then for $N\gg 1$ one finds:
$v_N \simeq v_0 - {\cal C}/\ln^2 N$, where ${\cal C}$ is a constant.

An analytic argument which explains this slow convergence, and allows
one to compute the coefficient ${\cal C}$, has been given by  Brunet
and Derrida \cite{BD}. Rather than reproducing the original derivation
from Ref. \cite{BD}, we prefer to present (directly for the case of
QCD) a qualitative argument \cite{IMM04} which explains the most
salient feature of their result, namely its slow convergence to the
mean field limit as $N\to\infty$.

This is related to the fact that, as mentioned before, the microscopic
front has a compact width, and therefore its evolution is frozen in a
state of `pre--asymptotic velocity' \cite{IMM04}. The {\em width} of
the front is the distance $\Delta\rho_{\rm f}=\rho -\rho_s$ over which
the amplitude $T_\tau(\rho)$ decreases from $T_\tau(\rho_s)\sim 1$ down
the minimal allowed value $T\sim\alpha_s^2$. This can be estimated by
using the mean field expression (\ref{Trhos}) for the amplitude.
Indeed, the MFA becomes appropriate (for a single front) as soon as
$T\gg \alpha_s^2$, that is, everywhere except in the few bins nearby
the tip of the front. But since the front is relatively wide (see
below), one can neglect the tip region when estimating its width. Then
Eq.~(\ref{Trhos}) implies $\Delta\rho_{\rm f}\sim (1/\gamma_s)\ln
(1/\alpha_s^2)$, which is large indeed.

Now, from the discussion after Eq.~(\ref{Trhos}), we know that the
front sets in diffusively, and thus requires a formation `time' :
Eq.~(\ref{Trho}) shows that, for the front to spread over a given
distance $\rho -\rho_s$, it takes a rapidity evolution
 \be
 \bar\alpha_s \,\tau\sim
 \,\frac{(\rho-\rho_s)^2}{2\beta}\,.
 \label{diffusiont}
 \ee
Through this evolution, the velocity of the front increases towards its
asymptotic value according to Eq.~(\ref{lsdef}). If the front is
allowed to extend arbitrarily far away, as it was the case for the MFA,
then the velocity will asymptotically approach the value $\bar\lambda$.
However, when the front is compact, as for the discrete system, the
formation time is finite as well, namely of the order
 \be\label{Dtau}
 \bar\alpha_s \Delta \tau
 \sim \,\frac{(\Delta\rho_{\rm f})^2}{2\beta}\,
 \sim\,
 \frac{\ln^2 (1/\alpha_s^2)}{2 \beta\,\gamma_s^2}\,,\ee
which implies that the front velocity cannot increase beyond a value
\be\label{ls}
 \lambda_{\rm as}\simeq \bar\lambda_{\rm as} -\kappa\,\bar\alpha_s\,
\frac{\gamma_s\beta}{\ln^2(1/\alpha_s^2)}\,. \label{satscal}
 \ee
This estimate is valid when $\alpha_s^2\ll 1$. The fudge factor
$\kappa$ cannot be determined by this qualitative argument, but this is
computed in Refs. \cite{BD,MS04} as $\kappa=\pi^2/2$.

The first term in \E{ls} is the mean field estimate $\bar\lambda_{\rm
as}\simeq 4.88 \bar\alpha_s$. But the second, corrective, term is
particularly large, not only because it decreases very slowly with
$\alpha_s^2$, but also because its coefficient is numerically large:
$\pi^2 \gamma_s \beta/2\approx 150$. Thus, although Eq.~(\ref{ls})
becomes an {\em exact result} when $\alpha_s^2$ is arbitrarily small,
this result remains useless for practical applications.

 Let us finally notice that, because of the compact nature of
the front, and thus of corresponding formation time, the asymptotic
velocity (\ref{ls}) is reached {\it exponentially} fast in $\tau$, with
a typical relaxation `time' given by Eq.~(\ref{Dtau}): $\bar\alpha_s
\Delta \tau \sim \ln^2 (1/\alpha_s^2)$ \cite{Panja}. This feature too
is at variance with the MFA, where the corresponding approach
is only power--like, cf. Eq.~(\ref{lsdef}).

\section{The breakdown of geometric scaling}
\label{SECT_DIFF}
\setcounter{equation}{0}

In the previous subsection, we have followed a particular realization
of the evolution, represented by a front
$T_\tau(\rho)$, and we have computed the saturation exponent as the
velocity of this front. Since the evolution is stochastic, different
realizations of the same evolution will lead to an {\it ensemble} of
fronts, which corresponds to the ensemble of configurations introduced
in Sect. \ref{SECT_FLUCT}. The precedent discussion applies to any of
these fronts: They all have the same asymptotic velocity, as given by
Eq.~(\ref{ls}), and except for the foremost region around the tip of
the front, they all have the same shape, namely the shape predicted by
the BK equation. However, in general these fronts will be displaced
with respect to each other along the $\rho$--axis, leading to a {\em
front dispersion}. That is, the position $\rho_s$ of the front is
itself a random variable, characterized by an expectation value
$\lan\rho_s\ran_\tau$, with
 \be
 \lan\rho_s\ran_\tau  \ {\simeq}\
  \lambda_{\rm as}\tau\,,\qquad{\rm for}\qquad
 \bar\alpha_s\tau \gg \ln^2 (1/\alpha_s^2)\,,
 \ee
but also by a dispersion $\sigma^2(\tau)\equiv \lan\rho_s^2\ran_\tau -
\langle\rho_s\rangle_\tau^{\,2}$.
Physically, this dispersion originates in the rare fluctuations
discussed in Sect. \ref{SECT_FLUCT} : In a particular realization, a
dipole with unusually small size may be created, which after further
evolution will ``pull'' the whole front behind him far ahead of the
typical evolution, resulting in a saturation scale $\rho_s(\tau)$ for
this particular realization which is larger than the mean value
$\langle\rho_s\rangle_\tau$.

To study the evolution of the ensemble of fronts in QCD, we shall rely
again on the corresponding studies in statistical physics. These
studies show that the position $\rho_s$ of the front executes a random
walk around its average value, so that the front dispersion rises
linearly with $\tau$ :
 \be
 \sigma^2(\tau)\ \simeq\ D_{\rm fr}\,\bar\alpha_s\tau\,,\ee
where $D_{\rm fr}$ is known as the {\it front diffusion coefficient}.
Besides, the numerical studies, which for some models have been pushed
up to astronomically large values of $N$ (as large as $10^{160}$),
demonstrate that $D_{\rm fr}$ scales like $1/\ln^3 N$ when $N\gg 1$
\cite{BD,Moro041}. So far, this is a purely numerical observation, for
which there is no fundamental understanding (see however \cite{Panja}).
Translating this result to QCD, one finds \cite{IMM04}:
 \be\label{Dfr} D_{\rm fr}\ \simeq\
 \frac{\cal D}{\ln^3(1/\alpha_s^2)}
 \qquad {\rm when}\qquad \alpha_s\,\ll\, 1\,,\ee
with an unknown coefficient ${\cal D}$. When decreasing $\alpha_s$,
$D_{\rm fr}$ vanishes very slowly.

\begin{figure}
  \includegraphics[width=.95\textwidth]{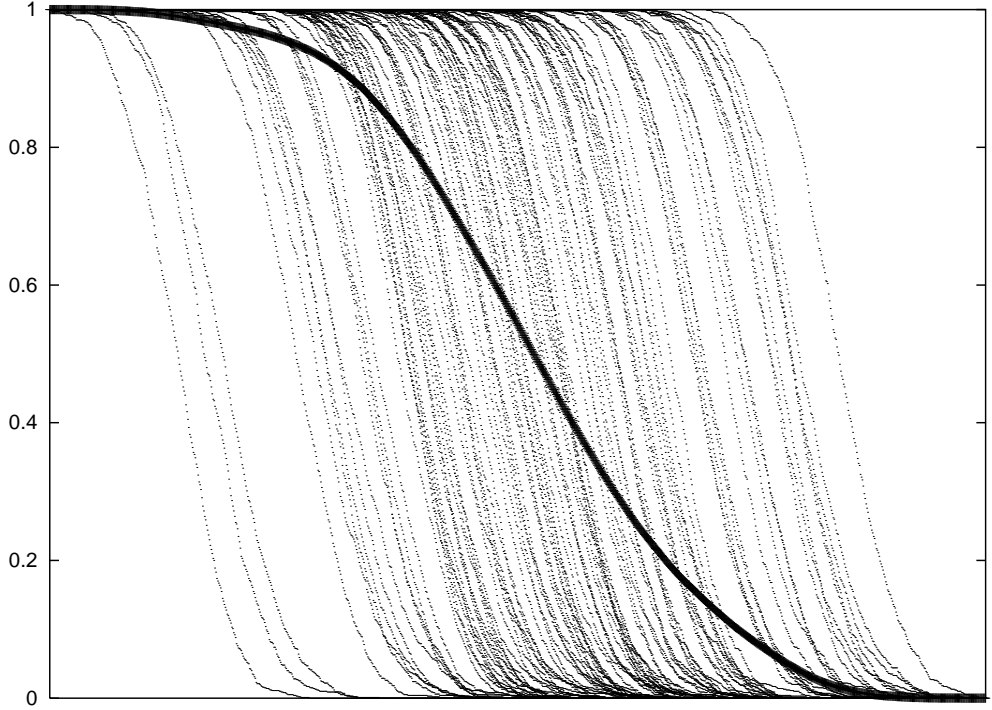}
  \caption{The scattering amplitude
$T$ for different partonic realizations at a given rapidity against
$\rho=\ln(r_0^2/r^2)$. The thick line is the average over all
realizations, i.e. the physical amplitude $\lan T(\rho) \ran$, see
Eq.~(\ref{average}).} \label{WANDER}
\end{figure}

\begin{figure}[t]    
    \centerline{\epsfxsize=7.cm\epsfbox{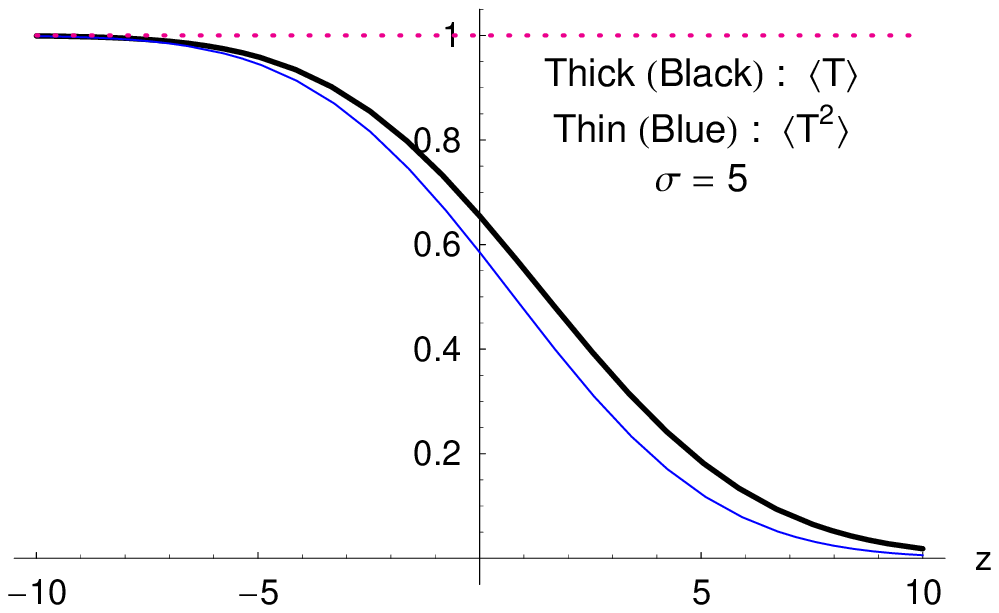}
    \hspace{0.25cm}
    \epsfxsize=7.cm\epsfbox{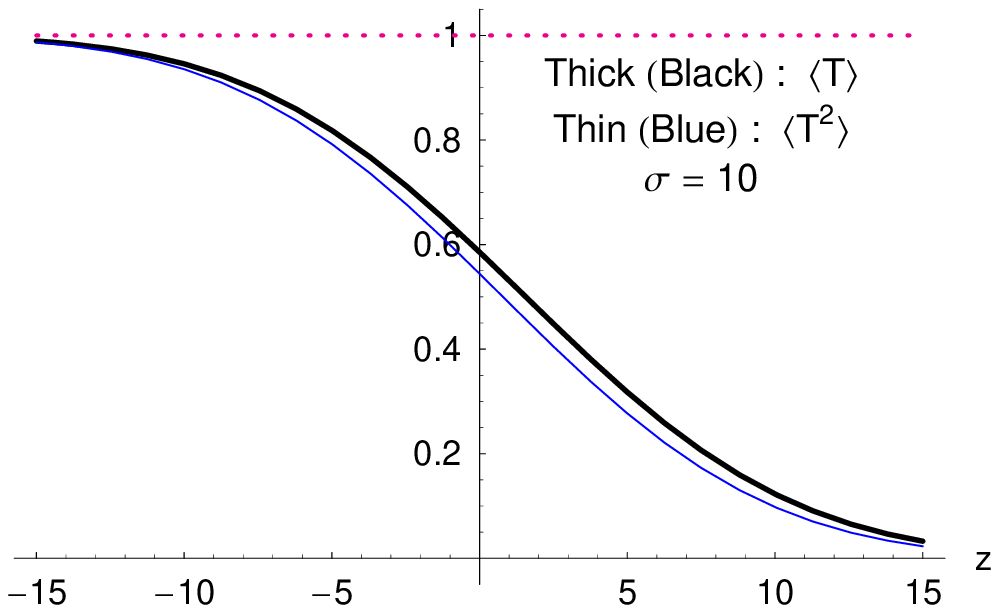}}
    \caption{\sl Evolution of $\langle T \rangle_\tau$ and
                 $\langle T^{2} \rangle_\tau$ with increasing $\sigma$.
                 \label{sigma}}
                 \vspace*{0.5cm}
\end{figure}

This diffusive wandering of the fronts is illustrated in
Fig.~\ref{WANDER} \cite{IMM04}. (This applies to the discrete
statistical model in Ref.~\cite{BD}, but a similar situation is
expected also in QCD.) All the fronts represented here are different
realizations of the same evolution; that is, they have been obtained by
evolving the same initial condition over the same period of time.  The
dispersion of $\rho_s$ is manifest in this picture, and so is also the
universality of the shape of the individual fronts for $T$. However,
precisely because of the dispersion, the shape of the {\em average}
amplitude $\lan T(\rho)\ran_\tau$ (represented by the thick line in
Fig.~\ref{WANDER}) is quite different from the shape of the individual
fronts, and, besides, this average shape is changing with $\tau$ : when
increasing $\tau$, the dispersion increases, so the average front $\lan
T(\rho)\ran_\tau$ becomes flatter and flatter, as visible on
Fig.~\ref{sigma}. We conclude that, as a consequence of fluctuations,
{\em geometric scaling is violated for the average amplitude}
\cite{MS04,IMM04}.

To study this violation in more detail, notice that the average
amplitude can be computed as an {\em average over $\rho_s$} :
 \begin{equation}\label{average}
    \lan T(\rho) \ran_\tau =
    \int \limits_{-\infty}^{\infty}
    d\rho_s\, P(\rho_s,\tau)\, T(\rho,\rho_s),
\end{equation}
where $P(\rho_s,\tau)$ is a Gaussian probability distribution for
$\rho_s$ :
  \begin{equation}\label{probdens}
    P(\rho_s,\tau) =
    \frac{1}{\sqrt{\pi}\,\sigma(\tau)}\,
    \exp \left[
    -\frac{\left( \rho_s - \langle \rho_s \rangle_\tau
   \right)^2}{\sigma^2(\tau)}
    \right],
\end{equation}
and $T(\rho,\rho_s)$ is the shape of the individual fronts, cf.
Eq.~(\ref{Trhos}), and can be schematically written as :
 \begin{equation}\label{Tevent}
    T(\rho,\rho_s)=
    \begin{cases}
        \displaystyle{1} &
        \text{ for\,  $\rho \leq \rho_s$}
        \\
        \displaystyle{\exp \left[ -\gamma_s (\rho - \rho_s) \right]} &
        \text{ for\,  $\rho \geq \rho_s$}.
    \end{cases}
\end{equation}
The behaviour of the average amplitude as a function of the `scaling
variable' $z\equiv \rho - \lan \rho_s \ran_\tau $ depends upon the
competition between $\sigma$ (the width of the Gaussian distribution of
the fronts) and $1/\gamma_s \sim 1$, which characterizes the
exponential decay of the individual fronts.  We can thus distinguish
between two types of behaviour, one at intermediate energies, the other
one at high energies:

\bigskip
 \noindent $\bullet$ \underline{$\sigma \ll 1$}
\bigskip

Since $\sigma \simeq \sqrt{D_{\rm fr}\bar\alpha_s\tau}$ where $D_{\rm
fr}$ vanishes when $\alpha_s^2\to 0$, one can find a regime where the
diffusion of the front plays no role, and geometric scaling still
holds: This happens for not too large $\bar\alpha_s\tau$ and
sufficiently small $\alpha_s$, such that $\sigma\ll 1$. Then one finds:
 \be\label{Tsmallt}
 \lan T(\rho) \ran_\tau \simeq \exp(-\gamma_s z)\qquad {\rm for}
 \qquad z\gg \sigma\,.\ee
That is, for a given (small) $\alpha_s^2$, and over a limited evolution
in rapidity, the average amplitude retains the shape of the individual
fronts.

\bigskip \noindent $\bullet$ \underline{$\sigma \gg 1$}
\bigskip

However, the typical situation at high energy is such that $\sigma\gg
1$. In that case, and for all values of $z$ such that $z \ll \sigma^2$,
one finds that $\lan T \ran$ is dominated by the saturating piece $T=1$
of Eq.~(\ref{Tevent}), and thus is {\em completely insensitive} to the
BFKL profile of the individual fronts. Namely:
 \begin{equation}\label{Thighsigma}
    \lan T \ran_\tau \,\simeq\,
    \frac{1}{2}\, {\rm Erfc}\left(\frac{z}{\sigma} \right)
    \qquad {\rm for} \quad -\infty < z \ll \sigma^2,
\end{equation}
where ${\rm Erfc(x)}$ is the complimentary error function, for which we
recall that
\begin{equation}\label{erfc}
    {\rm Erfc}(x)=
    \begin{cases}
        \displaystyle{2-\frac{\exp(-x^2)}{\sqrt{\pi}x}} &
        \text{ for\,  $x \ll -1$}
        \\*[0.1cm]
        \displaystyle{1} &
        \text{ for\,  $x=0$}
        \\*[0.1cm]
        \displaystyle{\frac{\exp(-x^2)}{\sqrt{\pi}x}} &
        \text{ for\,  $x \gg 1$}.
    \end{cases}
\end{equation}
Eq.~(\ref{Thighsigma}) shows that, in this high--energy regime, the
average amplitude scales as a function of $z/\sigma$, that is
\cite{IMM04}
 \begin{equation}
 \lan T(\rho) \ran_\tau \simeq {\cal T}
 \left(\frac{\rho-\langle\rho_s\rangle_\tau}{\sqrt{\bar\alpha_s
 \tau/\ln^3(1/\alpha_s^2)}}\right)\ , \label{scaling}
 \end{equation}
which is however a {\em different type of scaling} as compared to the
geometric scaling (compare to Eq.~(\ref{Trhos})): With increasing
$\tau$, and as a function of $z$, $\lan T(z) \ran_\tau$ becomes flatter
and flatter, as illustrated in Fig.~\ref{sigma} \cite{IT04}.
\begin{figure}[t]
    \centerline{\epsfxsize=7.cm\epsfbox{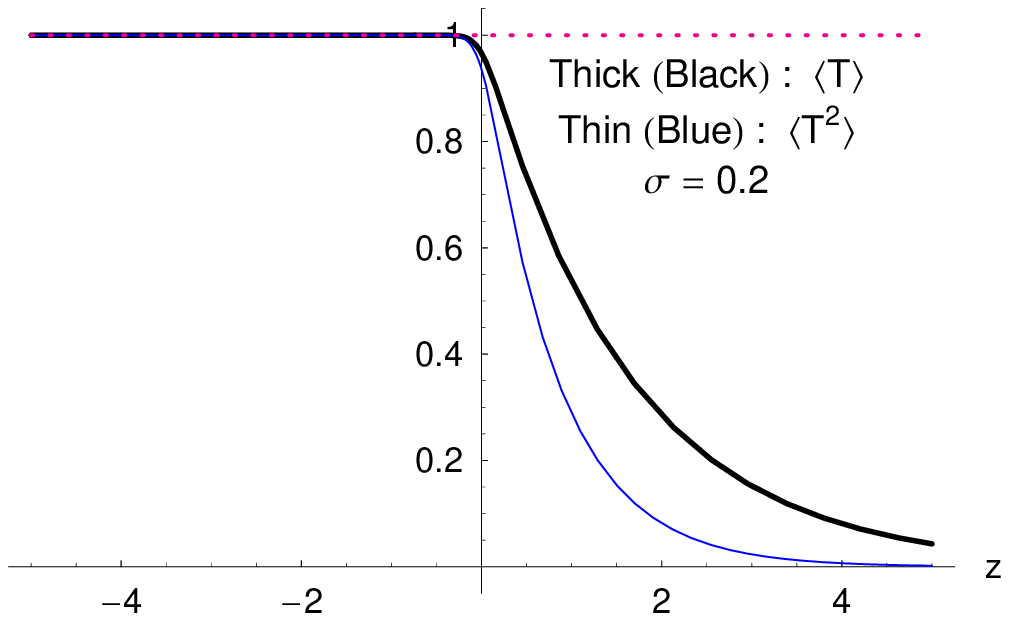}
    \hspace{0.25cm}
    \epsfxsize=7.cm\epsfbox{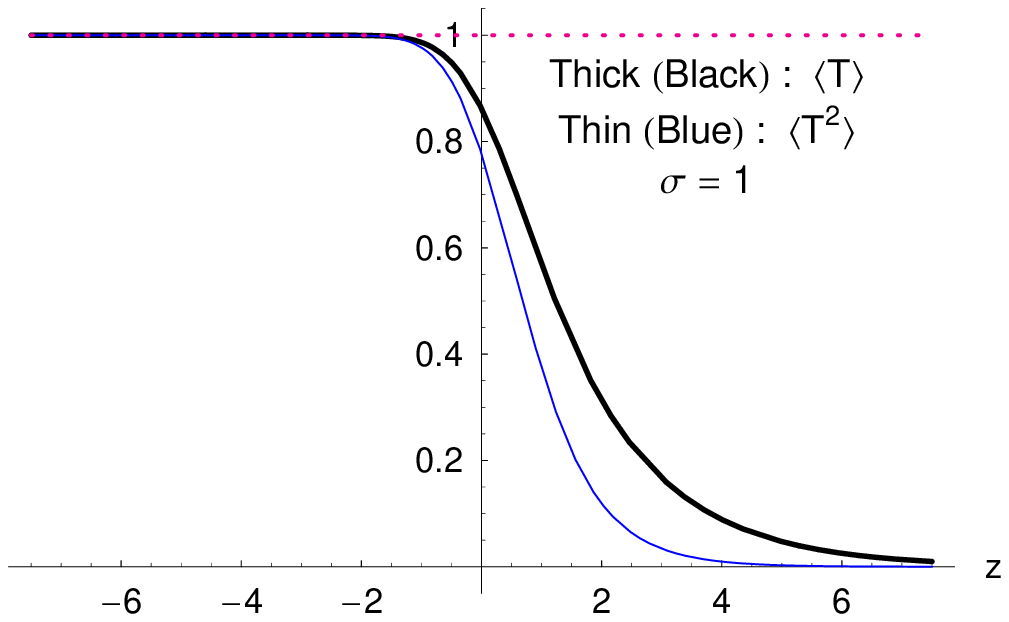}}
\end{figure}

The estimate (\ref{Thighsigma}) holds, in particular, in the range
$\sigma \ll z \ll \sigma^2$ where $\lan T \ran_\tau$ is small, $\lan T
\ran_\tau\ll 1$, yet is very different from the corresponding BFKL
prediction (compare in this respect the expression in the last line in
Eq.~(\ref{erfc}) to the BFKL amplitude (\ref{Trho})). To better
emphasize how dramatic is the breakdown of the BFKL approximation, let
us also compute the $n$--point correlation functions $\lan T^{(n)}
\ran_\tau$ (at equal points, for simplicity). One can obtain $\lan T^n
\ran_\tau$ by simply replacing $\gamma_s \rightarrow n\gamma_s$ in the
previous formulae.  Then one immediately finds that, in the
high--energy regime where $\sigma \gg 1$, {\em all} the higher
correlations $\lan T^n \ran_\tau$ are given by the {\em same}
expression, namely by Eq.~(\ref{Thighsigma}) (see also
Fig.~\ref{sigma}) \cite{IT04}
\begin{equation}\label{TnequalTn}
    \lan T^n \ran_\tau \simeq
    \lan T \ran_\tau
    \qquad {\rm for} \quad -\infty < z \ll\sigma^2.
\end{equation}
This signals a total breakdown of the mean field approximation, except
in the saturation regime where $\lan T \ran_\tau\simeq 1$ (see also
Fig.~\ref{sigma}). This is so because, in the presence of fluctuations
and for sufficiently large $\tau$, average quantities like $\lan
T^{(n)}\ran_\tau$ are dominated by those fronts within the statistical
ensemble which are at saturation for the value of $\rho$ of interest,
and this even when $\rho$ is well above the average saturation momentum
$\lan\rho_s\ran_\tau$, so that the corresponding {\it average}
amplitude is small.

\section{Pomeron loops}
\setcounter{equation}{0}

So far, our discussion has been mostly qualitative, and the language
used was essentially that of statistical physics. But it is also
interesting to understand these results within the more traditional
language of perturbative QCD, that is, in terms of Feynman graphs and
evolution equations. This is especially important in view of the
limitations of the correspondence with the statistical physics, which
so far has only allowed us to obtain asymptotic results
(valid when $\bar\alpha_s\tau \to\infty$ and $\alpha_s^2\to 0$) like \E{ls}.
To go beyond these results, we need the actual evolution equations in
QCD in the presence of both fluctuations and saturation. These
equations have been constructed in the large--$N_c$ limit
\cite{IT04,MSW05,IT05}, by combining the Balitsky equations
(or the CGC formalism) in the high density regime
with the dipole picture in the dilute regime. To motivate the structure
of these equations, we shall first discuss the diagrammatic interpretation
of the particle number fluctuations.

We shall use, as before, the dipole picture for the target wavefunction
in the dilute regime. Then, fluctuations in the dipole
number appear because of the possibility that one dipole internal
to the target splits into two
dipoles in one step of the evolution. In the discussion of
\E{B1} we have already shown, in Fig. \ref{FIG_1DIP}, the
basic diagram for dipole splitting. In that discussion, the dipole
appeared as the {\em projectile}, and the evolution was viewed as {\em
projectile evolution} (that is, the small rapidity increment $d\tau$
was given to the projectile). Here, we would like to visualize the
relevant fluctuations as splittings of the elementary dipoles inside
the target, and to that aim we need to perform {\em target
evolution}.

\begin{figure}[t]
    \centerline{\epsfxsize=12cm\epsfbox{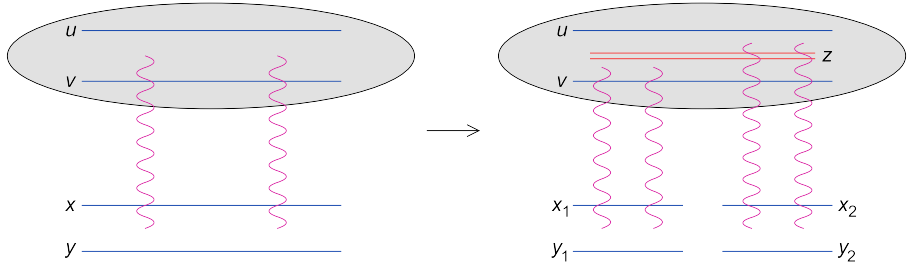}}
    \caption{\sl Diagrammatic illustration of the fluctuation
term in Eq.~(\ref{T2evol}) : the original dipole $(\bm{u},\bm{v})$
within the target splits at the time of the interaction into two new
dipoles $(\bm{u},\bm{z})$ and $(\bm{z},\bm{v})$, which then scatter off
two external dipoles. \label{Fig_split}}\vspace*{.5cm}
\end{figure}

In Fig. \ref{Fig_split} we show one
step in the evolution of the target, in which one of the dipoles
there --- the one with legs at $\bm{u}$ and  $\bm{v}$ ---
has split into two new dipoles (with coordinates
$(\bm{u},\bm{z})$ and $(\bm{z},\bm{v})$, respectively).
As further illustrated there, the original dipole can be probed via
scattering with {\em one} external dipole $(\x,\y)$,
in which case it provides a
contribution to the  scattering amplitude $\langle T(\x,\y)
\rangle_\tau$ at the original rapidity $\tau$. After evolution, the two
child dipoles can be measured via the scattering  with {\em two}
external dipoles, thus giving a contribution to the respective
amplitude $\langle T^{(2)}(\bm{x}_1,\bm{y}_1;\bm{x}_2,\bm{y}_2)
\rangle_{\tau+d\tau}$ at rapidity $\tau+d\tau$. This is in agreement
with the discussion in Sect. \ref{SECT_FLUCT} where we have seen that
one needs to scatter two external dipoles in order to be sensitive to
fluctuations.

From the previous discussion, one can also understand what should be
the role of the process in Fig. \ref{Fig_split} in the evolution of the
scattering amplitudes for external dipoles: This process generates a
change in the two--dipole scattering amplitude $\langle
T^{(2)}\rangle_\tau$ which is proportional to single--dipole amplitude
$\langle T \rangle_\tau$. Specifically, the following evolution
equation can be written down by inspection of Fig. \ref{Fig_split}
\cite{IT05}:
 \begin{align}\label{T2evol}
    \frac{\partial \mean{T^{(2)}(\bm{x}_1,\bm{y}_1;\bm{x}_2,\bm{y}_2)}}
    {\partial \tau}
    \bigg|_{\rm fluct}\!\!\! =
    \left(\frac{\alpha_s}{2\pi}\right)^2
    \frac{\abar}{2 \pi}\!
    \int\limits_{\bm{u},\bm{v},\bm{z}}&
    \mathcal{M}(\bm{u},\bm{v},\bm{z})\,
    \mathcal{A}_0(\bm{x}_1,\bm{y}_1|\bm{u},\bm{z})\,
    \mathcal{A}_0(\bm{x}_2,\bm{y}_2|\bm{z},\bm{v})\,
    \nonumber \\
    &\times\nabla_{\bm{u}}^2 \nabla_{\bm{v}}^2\, \mean{T(\bm{u},\bm{v})}.
\end{align}
The r.h.s. of this equation should be read as follows: A dipole
$(\bm{u},\bm{v})$ from the target splits into two new dipoles
$(\bm{u},\bm{z})$ and $(\bm{z},\bm{v})$ with probability $(\abar/2\pi)
\mathcal{M}(\bm{u},\bm{v},\bm{z})$ (cf. \E{dPsplit}), then the two
child dipoles scatter off the external dipoles, with an amplitude
$\alpha_s^2\mathcal{A}_0$ for each scattering. Finally, the
`amputated' amplitude $(1/\alpha_s^2)\nabla_{\bm{u}}^2
\nabla_{\bm{v}}^2\, \mean{T(\bm{u},\bm{v})}$ is, up to a normalization
factor, the dipole density $\mean{n(\bm{u},\bm{v})}$ in the target, as
obtained by inverting \E{Tconv}.

As indicated in the l.h.s. of \E{T2evol}, this equation describes only
that contribution to the evolution of  $\langle T^{(2)}\rangle_\tau$
which is associated with fluctuations in the dipole number. In addition
to that, there are standard terms describing the BFKL evolution and the
unitarity corrections \cite{B,RGE}.

\begin{figure}[t]
    \centerline{\epsfxsize=12cm\epsfbox{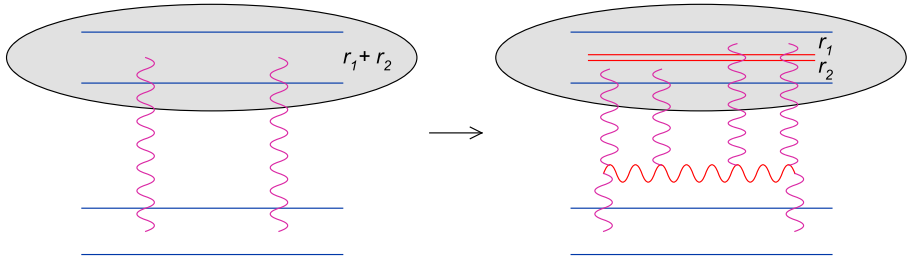}}
    \caption{\sl Two steps in the evolution of the average
scattering amplitude of a single dipole: the original amplitude (left)
and the Pomeron loop generated after two steps (right).
 \label{Two_steps}}
\end{figure}

Let us finally consider the evolution of the dipole scattering
amplitude $\langle T \rangle$ after {\em two} steps. This involves
several processes, but the most interesting among them is the one
displayed in Fig. \ref{Two_steps}, which is sensitive to both
fluctuations and saturation. Specifically, the first step of the
evolution is the same as in Fig. \ref{Fig_split}: one dipole in the
target wavefunction splits into two, which implies that the original
$\langle T \rangle$ evolves into a $\langle T^{(2)}\rangle$ (cf.
\E{T2evol}). In the second step, the $\langle T^{(2)}\rangle$ evolves
back into a $\langle T \rangle$, according to the non--linear term in
\E{B1}. The latter process has been already represented from the
perspective of projectile evolution in Fig. \ref{FIG_1DIP}.d. In the
lower half part of Fig. \ref{Two_steps}, this process is now
represented as target evolution: From this perspective, it describes
the merging of four gluons into two.

Altogether, the two--step evolution depicted in Fig. \ref{Two_steps}
generates the simplest {\em Pomeron loop}, where by `Pomeron' we mean
at this level the two--gluon exchange between two dipoles (but
subsequent evolution will turn such two--gluon exchanges into BFKL
pomerons). The loop is constructed with two `triple--Pomeron vertices'
(the dipole kernels $\bar\alpha_s \mathcal{M}$) and two `Pomeron
propagators' (the dipole--dipole scattering amplitudes
$\alpha_s^2\mathcal{A}_0$). It turns out that these are the same
vertices as those describing the merging and splitting of {\em BFKL
Pomerons} within perturbative QCD \cite{BW93,BV99,BLV05}. A simple
effective theory for Pomeron dynamics which combines all these
ingredients --- the BFKL evolution of the Pomerons together with their
interactions: dissociation (one Pomeron splitting into two) and
recombination (two Pomerons merging into one) --- and reproduces the
correct evolution equations in QCD at large $N_c$
\cite{IT04,MSW05,IT05} has been proposed in Ref. \cite{BIIT05}. In
particular, in Ref. \cite{BIIT05} one can find an explicit expression
for the Pomeron loop in Fig. \ref{Two_steps}.

In Sect. \ref{SECT_EBE}, we have mentioned that the BK equation ---
which, we recall, is a mean field approximation to the QCD evolution at
high energy --- is in the same universality class as the FKPP equation
of statistical physics \cite{MP03}. It is interesting to mention at
this point that the complete evolution in QCD at large $N_c$, which
includes the effects of particle number fluctuations via terms like
\E{dPsplit}, is in the same universality class as the {\em stochastic
FKPP equation} (sFKPP) \cite{IT04} --- a Langevin equation with a
specific `noise term' which simulates particle number fluctuations
\cite{Panja}. By further studying this equation (in particular, via
numerical calculations), one should be able to go beyond the asymptotic
results presented in Sects. \ref{SECT_EBE} and \ref{SECT_DIFF}, and
find the behaviour of the scattering amplitudes in QCD for realistic
values of the energy and of the coupling constant. This program is
currently under way.


\end{document}